\documentclass[11pt]{article}
\usepackage{amssymb,amsthm,fullpage}

\newtheorem{definition}{Definition}
\newtheorem{lemma}{Lemma}
\newtheorem{theorem}{Theorem}
\newtheorem{corollary}{Corollary}
\newtheorem{observation}{Observation}

\title{On Low Distortion Embeddings of Statistical Distance Measures into Low Dimensional Spaces\footnote{The short version of this paper was accepted for presentation at the 20$^{th}\!$ International Conference on Database and Expert Systems Applications, DEXA 2009.}}
\author{Arnab Bhattacharya \and Purushottam Kar \and Manjish Pal}
\date{Department of Computer Science and Engineering\\Indian Institute of Technology Kanpur, INDIA.\\\vspace{0.1in}\texttt{\{arnabb,purushot,manjish\}@cse.iitk.ac.in}}

\begin{document}

\maketitle

\begin{abstract}
Statistical distance measures have found wide applicability in information
retrieval tasks that typically involve high dimensional datasets. In order
to reduce the storage space and ensure efficient performance of queries,
dimensionality reduction while preserving the inter-point similarity is
highly desirable. In this paper, we investigate various statistical distance
measures from the point of view of discovering low distortion embeddings
into low-dimensional spaces. More specifically, we consider the Mahalanobis
distance measure, the Bhattacharyya class of divergences and the
Kullback-Leibler divergence. We present a dimensionality reduction method
based on the Johnson-Lindenstrauss Lemma for the Mahalanobis measure that
achieves arbitrarily low distortion. By using the Johnson-Lindenstrauss
Lemma again, we further demonstrate that the Bhattacharyya distance admits
dimensionality reduction with arbitrarily low additive error. We also
examine the question of embeddability into metric spaces for these distance measures due to the
availability of efficient indexing schemes on metric spaces. We provide
explicit constructions of point sets under the Bhattacharyya and the
Kullback-Leibler divergences whose embeddings into any metric space incur
arbitrarily large distortions. We show that the lower bound presented for
Bhattacharyya distance is nearly tight by providing an embedding that
approaches the lower bound for relatively small dimensional datasets.
\end{abstract}

\section{Introduction}
\label{intro}

The problem of embedding distance measures into normed spaces arises in
applications dealing with huge amounts of high-dimensional data where performing
point, range or nearest-neighbor (NN) queries in the ambient space entails
enormous computational costs. The prohibitive costs arise in most cases due to
two factors. First, calculating pairwise distances and answering
proximity queries such as $k$-NN queries becomes costlier as the dimensionality
of the data increases (a phenomenon more commonly known by the epithet
``curse of dimensionality''). Second, in many cases such as image
databases, the imposed distance measures such as the earth mover's distance \cite{rubner} are inherently expensive to compute. The problem of indexing and
searching is magnified if the distance measures do not form a metric.

Various approaches have been proposed to solve this problem. These include
obtaining easily estimable upper/lower-bounds on the distance measures
\cite{emd,rubner} and finding embeddings which allow specific proximity queries
to be efficiently solved \cite{lce,indyk}. These methods have been found to
be crucial for database retrieval algorithms in obtaining speedups over na\"ive
search techniques.  This is mainly due to the fact that, in many of these
examples, the lower/upper-bounds or the embeddings involve an $l_p$ metric for which there exist efficient algorithms for answering proximity queries
\cite{moti-kitab}.

The concept of a ``low distortion'' embedding (defined in Section~\ref{prelim})
is very useful in this context. A low distortion embedding ensures that notions
of distance are preserved almost intact. This is certainly most desirable since
it gives performance guarantees in terms of accuracy for all proximity queries
by preserving the geometry of the original space almost exactly as against many
of the methods cited above which are optimized for specific types of queries. The presence of several index structures for metric
spaces (especially for the Euclidean space) \cite{moti-kitab} makes it natural to
try to approximate a given distance measure by a metric distance (especially
Euclidean distance). Getting an approximation of the original distance measure
in terms of a metric distance also shows how inherently ``geometric'' is the
distance measure. In fact if the new metric distance is $l_2$, there is the
added benefit of being able to drastically reduce the dimensionality of the
objects in the database while incurring a very small distortion via the
Johnson-Lindenstrauss lemma \cite{jl-orig}. Apart from its practical importance,
such a result that quantifies the relation between the distance measures is of
theoretical interest as well, and results of this nature, especially when both
the distance measures are metrics, have gained huge interest in theoretical
computer science over the past few years \cite{jiri}. 

A more interesting, and often more difficult, situation arises in case of
statistical distance measures. They are widely used in database and pattern
recognition applications and typically involve high-dimensional data. It has
been found that in many scenarios, especially in similarity based search in
image retrieval \cite{similar-search}, statistical distance measures like the
Mahalanobis \cite{maha} and Bhattacharyya \cite{bhat} measures give better
performance than the standard $l_2$ distance. In the field of bioinformatics,
the Mahalanobis distance measure has been found to be more useful than the $l_2$ distance when the distance between two DNA sequences is measured by
simultaneously comparing the frequencies of all subsequences of $n$ adjacent
letters (i.e., $n$-words) in the two sequences~\cite{bio-maha}. Furthermore, the Mahalanobis distance has found application in face recognition as well \cite{ijcnn-maha}. The Bhattacharyya class of distance measures which include the Bhattacharyya distance and the Hellinger distance are also widely used in
diverse database scenarios such as nearest-neighbor classification
\cite{kbs-hell-1}, detecting voice over IP floods \cite{pds-hell}, and
recognition tasks \cite{cvpr-hell}. However, it should be noted that often
applications of the Bhattacharyya distance use the distance measure to calculate the dissimilarity between multivariate normal distributions; we do not address that in this paper. Another important and widely used statistical similarity measure is the Kullback-Leibler divergence \cite{kl}. It has been shown that the Kullback-Leibler divergence is well suited for use in
real-time image segmentation algorithms and time-critical texture classification and retrieval from large databases \cite{tex-sim}. It has also been used in machine learning to design pattern classifiers for high-dimensional image spaces \cite{kl-boost}. Apart from its applications this measure is interesting from a theoretical perspective as well because of its information-theoretic roots.

These distance measures have received a lot of attention recently and have been
examined from several perspectives including clustering \cite{km-info-cluster},
NN-retrieval \cite{nn-bregman}, computing Voronoi diagrams
\cite{voronoi-bregman}, and sketching \cite{gim-unsketch}. We examine these
distance measures from another interesting perspective---that of low distortion
embeddings into metric spaces and dimensionality reduction. The lack of inherent
``geometric'' properties make them harder candidates for low distortion and
low-dimensionality embeddings into metric spaces. To the best of our knowledge,
this is the first investigation into these distance measures from the point of
view of dimensionality reduction and embeddability into metric
spaces.\\

\noindent
\textbf{Our Contributions}: In this paper, we examine three statistical distance measures with the goal of obtaining low distortion, low-dimensional embeddings for them.
The paper is structured as follows. In Section~\ref{prelim}, we provide 
definitions and well known results that will be used in the subsequent
discussion. In Section~\ref{hellinger}, we consider the Bhattacharyya distance
and prove that there cannot exist low distortion embeddings for the
Bhattacharyya distance into a metric space. We develop a technique which shows
that the lower bound on the distortion gets larger as we include distributions
that are ``farther'' from the uniform distribution in the probability simplex.
We also provide an embedding into the $l_2^2$ space (which forms a metric in the positive orthant) that approaches the lower bound for small dimensions.

In Section~\ref{kl}, using the technique developed in the previous section, we
show a similar result for the Kullback-Leibler divergence. We also develop
another technique that gives better bounds on the distortion when the set of
distributions under consideration does not contain distributions that are ``far
away'' from the uniform distribution. The section ends with a result relating
the Kullback-Leibler divergence and the $l_2^2$ measure.
Finally, in Section~\ref{qfd}, we provide low distortion embeddings for the
family of Quadratic Form Distances. As a special case, we show that the
Mahalanobis metric also admits a low distortion embedding. We end the paper with some concluding remarks in Section~\ref{conclude}.

\section{Preliminaries}
\label{prelim}

We begin by defining a few preliminary concepts related to metric spaces and
embeddings.
\begin{definition}[Metric Space]
A pair $M=(X,\rho)$ where $X$ is a set and $\rho:X \times X \longrightarrow
\mathbb{R}^+ \cup \{0\}$ is called a metric space provided the distance
measure $\rho$ satisfies the properties of identity, symmetry and triangular
inequality.
\end{definition}
\begin{definition}[$D$-embedding and Distortion]
\label{distortion}
Given two metric spaces $(X,\rho)$ and $(Y,\sigma)$, a mapping $f:X
\longrightarrow Y$ is called a $D$-embedding where $D \geq 1$, if there
exists a number $r >0$ such that for all $x,y \in X$, \[ r\cdot \rho(x,y)
\leq \sigma\left( {f(x),f(y)} \right) \leq D\cdot r \cdot \rho(x,y) \] The
infimum of all numbers $D$ such that $f$ is a $D$-embedding is called the
\emph{distortion} of $f$.
\end{definition}
In general, the factor $r$ is intended to allow the distances to be scaled by
some constant factor and does not affect the definition of the embedding (see
\cite{jiri} for details). It is easy to see that this notion of distortion can
be naturally extended to non-metric spaces as well.

A classic result widely used in the field of metric embeddings is the
Johnson-Lindenstrauss Lemma \cite{jl-orig} which makes it possible for large
point sets in high-dimensional Euclidean spaces to be embedded into
low-dimensional Euclidean spaces with arbitrarily small distortion.

\begin{theorem}[Johnson-Lindenstrauss Lemma \cite{jiri}]
Let $X$ be an $n$-point set in a $d$-dimensional Euclidean space (i.e.
$(X,l_2) \subset \left(\mathbb{R}^d,l_2\right)$), and let $\epsilon \in
(0,1]$ be given. Then there exists a $(1+\epsilon)$-embedding of $X$ into
$(\mathbb{R}^k,l_2)$ where $k = O\left( {\epsilon^{-2}\log n} \right)$.
Furthermore, this embedding can be found out in randomized polynomial time.
\end{theorem}

The main idea being used here is that the length of a $d$-dimensional vector
when randomly projected onto a lower $k$-dimensional space is sharply
concentrated around its expected value. This allows us to show the existence of
a low distortion embedding as well as obtain a randomized algorithm to discover
the embedding via random projections. The original proof of Johnson and
Lindenstrauss \cite{jl-orig} has been greatly simplified by the algorithmic
proofs of Gupta-Dasgupta \cite{jl-elem}, Indyk-Motwani \cite{cod} and
Arriaga-Vempala \cite{vem}. However, all these techniques involve sampling from
continuous distributions which make them somewhat impractical for database
purposes where one would like to perform operations via simple SQL queries. This
problem was solved by a beautiful result due to Achlioptas \cite{achlioptas} who
showed that one can in fact use a projection matrix with each entry chosen
independently from the distribution $U\{-1,+1\}$. (In fact, as shown in
\cite{vem}, any distribution that is symmetric about the origin with the second
moment as unity and bounded higher even moments can be used.) This is most
suited to a database application where the random projection can be applied by
simply splitting the attribute set into two halves by first sampling and summing up the attributes in each set, and then taking the difference of the two sums, and finally repeating this as many times as is the dimensionality of the projected space. We now state the main result of Achlioptas~\cite{achlioptas} which assures that our algorithmic results are readily applicable to database situations as well.

\begin{lemma}[\cite{achlioptas}]
\label{rand-proj}
Let $R = (r_{ij})$ be a random $d \times k$ matrix, such that each entry
$r_{ij}$ is chosen independently according to $U\{+1,-1\}$. For any fixed unit
vector $u \in \mathbb{R}^d$, and any $\epsilon > 0$, let $u' =
\sqrt{\frac{d}{k}}\left( {R^Tu} \right)$. Then, $E\left[ {\|u'\|^2} \right] = 1
= \|u\|^2$ and $ \Pr\left[ (1-\epsilon)\|u\|^2 < \|u'\|^2 < (1+\epsilon)\|u\|^2
\right] \geq 1 - e^{\frac{-k}{2}\left(\frac{\epsilon^2}{2} -
\frac{\epsilon^3}{3}\right)} $
\end{lemma}

This establishes the Johnson-Lindenstrauss lemma. For a given value of
$\epsilon$, the dimensionality of the projected space (i.e. $O\left(
{\epsilon^{-2}\log n} \right)$) is chosen in order to ensure an inverse
polynomial error probability which facilitates the application of the union
bound to ensure that none of the pairwise distances in the $n$-point set is
distorted too much. The result also ensures that even the inner products are
preserved to an arbitrarily low additive error; this is characterized by the
following corollary.

\begin{corollary}[\cite{vem}]
\label{jl-dot}
Let $u,v$ be unit vectors in $\mathbb{R}^d$. Then, for
any $\epsilon > 0$, a random projection of these vectors to yield the vectors $u'$ and $v'$ respectively satisfies $
\Pr\left[ {u \cdot v - \epsilon \leq u' \cdot v' \leq u \cdot v + \epsilon}
\right] \geq 1 -
4e^{\frac{-k}{2}\left(\frac{\epsilon^2}{2}-\frac{\epsilon^3}{3}\right)} $
\end{corollary}
\begin{proof}
Apply Lemma~\ref{rand-proj} to the vectors $u$, $v$ and $u-v$. The result
follows from using simple facts concerning inner products.
\hfill{}
\end{proof}

We shall be using the properties of the embedding described above to obtain
low distortion embeddings for various statistical distance measures. To
facilitate the discussion we refer to the process of mapping high-dimensional
point sets to low-dimensional ones via random projections as \emph{JL-type
embeddings}. We now develop some of the terminology that will be used later. In the discussion below, we assume that the histograms to be normalized, i.e., they
correspond to probability distributions.
\begin{definition}[Representative vector]
Given a $d$-dimensional histogram $P = (p_1,\dots p_d)$, let $\sqrt{P}$
denote the unit vector $(\sqrt{p_1}, \dots ,\sqrt{p_d})$. We shall call this
the \emph{representative vector} of $P$. 
\end{definition}
\begin{definition}[$\alpha$-constrained histogram]
A histogram $P = (p_1,p_2, \dots p_d)$ is said to be $\alpha$-constrained if
$p_i \geq \frac{\alpha}{d}$ for $i = 1,2,\ldots,d$.
\end{definition}
This ensures that the $\alpha$-constrained histograms have a level of
``smoothness'' to them and are not extremely skewed. The following result holds
for all $\alpha$-constrained histograms.

\begin{observation}
Given two $\alpha$-constrained histograms $P$ and $Q$, the inner product between
the representative vectors is at least $\alpha$, i.e., $\left\langle{
\sqrt{P},\sqrt{Q} }\right\rangle \geq \alpha$.
\end{observation}

For convenience, we will denote $\frac{\alpha}{d}$ by $\beta$. A $d$-dimensional
$\beta$-constrained distribution will then imply a distribution that is $\alpha$-constrained with $\alpha = \beta \cdot d$. Since the histograms are normalized,
$\alpha$ must be less than $1$. In other words, $\beta \leq \frac{1}{d}$.
We next examine three statistical distance measures starting with the
Bhattacharyya class of distance measures.

\section{The Bhattacharyya Class of Distance Measures}
\label{hellinger}

In the field of pattern classification, more specifically Bayesian decision
theory, the Bhattacharyya bound is an upper-bound on the expected error rate of
a Bayesian decision process \cite{dhs}. For a binary classification task on a
feature $x$ with the two categories as $\omega_1$ and $\omega_2$ and the
likelihood parameters characterized by the distributions $p(x|\omega_i)$ for $i
= 1,2$, the Bhattacharyya bound on the error probability of a Bayesian classifier, ignoring the priors, is given by
\[
\Pr\{\mathrm{error}\} \leq \int{\sqrt{p(x|\omega_1)p(x|\omega_2)}\ dx}
\]

This is known as the \emph{Bhattacharyya coefficient} between the two
distributions. For two histograms $P = (p_1,p_2, \dots ,p_d)$ and $Q = (q_1,q_2,
\dots q_d)$ with $ \sum_{i=1}^d p_i = \sum_{i=1}^n q_i = 1$ and each $p_i, q_i \geq
0$, the \emph{Bhattacharyya coefficient} \cite{bhat} is described as
\[
BC(P,Q) = \sum\limits_{i=1}^n{\sqrt{p_iq_i}}
\]

Using this coefficient, two distance measures can be defined as follows. The
\emph{Bhattacharyya distance} \cite{bhat} is defined as follows
\[
BD(P,Q) = -\ln BC(P,Q)
\]
It is easy to see that this measure does not form a metric. Another distance
measure in this class, namely, the \emph{Hellinger distance} \cite{pds-hell} between
distributions is defined as
\[
H(P,Q) = 1 - BC(P,Q) =
\frac{1}{2}\left(\left\|\sqrt{P}-\sqrt{Q}\right\|\right)^2
\]
The fact that $H(P,Q)$ is the Euclidean distance between the points $\sqrt{P}$
and $\sqrt{Q}$ allows us to state the following theorem.

\begin{theorem}
\label{hellinger-euclid}
The Hellinger distance admits a low distortion dimensionality reduction.
\end{theorem}
\begin{proof}
Since the Hellinger	distance between the two histograms is the Euclidean distance between
their representative vectors, given a set of histograms, if we subject
the corresponding set of representative vectors to a JL-type embedding then Lemma~\ref{rand-proj}
ensures that the Euclidean distance between the embedded representative vectors is a $1 \pm \epsilon$
approximation of the Hellinger distance between the corresponding histograms with high probability.
\end{proof}

\subsection{Dimensionality Reduction for the Bhattacharyya Distance}

We now consider the possibility of extending this idea to the Bhattacharyya
distance as well which would provide us with dimensionality reduction for the
Bhattacharyya distance. Given a set of histograms under the Bhattacharyya
distance that are $\alpha$-constrained and an error parameter $\epsilon$, we
perform a JL-type embedding with the error parameter appropriately set. This
embedding constitutes a dimensionality reduction as we impose the Bhattacharyya
distance on the new space. The following theorem shows that this embedding only
incurs a small additive error.

\begin{theorem}
A JL-type embedding of a set of $\alpha$-constrained histograms under the
Bhattacharyya distance measure with the error parameter set to $\epsilon' =
\frac{\epsilon \cdot \alpha}{2}$ incurs only an additive error of
$\epsilon$, i.e., if $P,Q$ are the initial histograms transformed
respectively to $P',Q'$, then 
\[
BD(P,Q) - \epsilon \leq BD(P',Q') \leq
BD(P,Q) + \epsilon
\]
\end{theorem}
\begin{proof}
By Corollary~\ref{jl-dot}, we have the following with high probability:
\[
\left\langle{ \sqrt{P},\sqrt{Q} }\right\rangle - \epsilon' \leq
\left\langle{ \sqrt{P'},\sqrt{Q'} }\right\rangle \leq \left\langle{
\sqrt{P},\sqrt{Q} }\right\rangle + \epsilon'
\]
Taking $-\ln()$ throughout
and using the definition of the Bhattacharyya distance, we have 
\begin{eqnarray*}
BD(P',Q') &\geq& BD(P,Q) - \ln\left({ 1 + \frac{\epsilon'}{\left\langle{
\sqrt{P},\sqrt{Q} }\right\rangle} }\right)\\
BD(P',Q') &\leq& BD(P,Q) + \ln\left({ \frac{1}{1 -
\frac{\epsilon'}{\left\langle{ \sqrt{P},\sqrt{Q} }\right\rangle}} }\right)
\end{eqnarray*}
Since the distributions are $\alpha$-constrained, we have $\left\langle{
\sqrt{P},\sqrt{Q} }\right\rangle \geq \alpha$. Hence, 
\begin{eqnarray*}
BD(P',Q') &\geq& BD(P,Q) - \ln\left({ 1 + \frac{\epsilon'}{\alpha}
}\right)\\
BD(P',Q') &\leq& BD(P,Q) + \ln\left({ \frac{1}{1 - \frac{\epsilon'}{\alpha}}}\right)
\end{eqnarray*}
For any $x$, $e^x \geq 1 + x$. Hence,
\[
\ln\left({ 1 +
\frac{\epsilon'}{\alpha} }\right) \leq \frac{\epsilon'}{\alpha}
\]
Also, the function $f(x) = 2x - \ln\left({ \frac{1}{1-x} }\right)$ is positive for all $x \leq \frac{1}{2}$. Hence, for $\frac{\epsilon'}{\alpha} \leq
\frac{1}{2}$ (which is true since $\epsilon' = \frac{\epsilon \cdot
\alpha}{2}$ and $\epsilon \leq 1$), we have
\[
\ln\left({ \frac{1}{1 -
\frac{\epsilon'}{\alpha}} }\right) \leq \frac{2\epsilon'}{\alpha}
\]
This implies the following bounds on $BD(P,Q)$
\[
BD(P,Q) - \frac{\epsilon'}{\alpha} \leq BD(P',Q') \leq BD(P,Q) +
\frac{2\epsilon'}{\alpha}
\]
which gives us the desired result since $\epsilon' = \frac{\epsilon \cdot \alpha}{2}$.
\end{proof}

The natural question that arises now is whether the Bhattacharyya distance,
being a non-metric, also admits low distortion embeddings into metric spaces.
Next, we develop a proof technique that shows that the distortion incurred by
any embedding of point sets under the Bhattacharyya distance into a metric space
can be made arbitrarily large by including appropriately chosen histograms.

\subsection{The Relaxed Triangle Inequality Technique}
\label{rti}

In order to present the proof, we first define \emph{$\lambda$-relaxed triangle
inequality} for a distance measure. This essentially parallels the definition of
a \emph{relaxed metric} as defined in \cite{km-info-cluster}.
\begin{definition}[$\lambda$-Relaxed Triangle Inequality]
A set $X$ equipped with a distance function $d:X \times X \longrightarrow
\mathbb{R}^+\cup\{0\}$, is said to satisfy the \emph{$\lambda$-relaxed triangle
inequality} if there exists some constant $\lambda \leq 1$ such that for all
triplets $p,q,r \in X$, the following holds
\[
d(p,r) + d(r,q) \geq \lambda \cdot d(p,q)
\]
\end{definition}
Metrics satisfy the $\lambda$-relaxed triangle inequality
for $\lambda = 1$. The following result allows us to arrive at lower bounds on the distortion of
embeddings into metric spaces for a distance measure that violates a relaxed
triangle inequality.
\begin{lemma}
\label{lb}
Any embedding of a set $X$ equipped with a distance function $d$ that does
not satisfy the $\lambda$-relaxed triangle inequality into a metric space
incurs a distortion of at least $\frac{1}{\lambda}$.
\end{lemma}
\begin{proof}
Since $(X,d)$ does not satisfy the $\lambda$-relaxed triangle inequality,
there exist points $p,q,s \in X$ such that $d(p,s) + d(s,q) < \lambda \cdot
d(p,q)$.  Now, let $(X,d)$ be embedded into a metric space $(Y, \rho)$ via
the mapping $f$ that incurs a distortion $D$. This implies that, for all
points $x,y \in X$, we have 
\[
r\cdot d(x,y) \leq \rho\left( {f(x),f(y)} \right) \leq D \cdot r \cdot
d(x,y)
\]
Next, consider the three points $p,q,s \in X$. Since $(Y, \rho)$ is a metric
space, it satisfies the triangle inequality for the embeddings of these points. Hence 
\begin{eqnarray*}
\rho(f(p),f(s)) + \rho(f(s),f(q)) &\geq& \rho(f(p),f(q))\\
rD \cdot d(p,s) + rD \cdot d(s,q) &>& r \cdot d(p,q)\\
rD\lambda \cdot d(p,q) &>& r \cdot d(p,q)
\end{eqnarray*} 
Hence we have $D > \frac{1}{\lambda}$.
\end{proof}
In general, if $(Y,\rho)$, i.e. the space being embedded into, satisfies a $\lambda'$-relaxed triangle inequality, then we get a lower bound of $\frac{\lambda'}{\lambda}$ on the value of
distortion of $(X,d)$ into $(Y,\rho)$.

\subsection{Lower Bound on Distortion for Embeddings into Metric Spaces}
\label{lb-bhat}

We now appeal to the relaxed triangle inequality argument by constructing point
sets under the Bhattacharyya distance that fail to satisfy the relaxed triangle inequality and then applying Lemma~\ref{lb} to get a lower bound on the distortion.

The idea is to choose three distributions $P,Q,R$ such that the angle between
the representative vectors of $P$ and $R$ is almost $\pi/2$ (i.e., the
similarity is almost $0$) while the angle between the representative vectors of
$P$ and $Q$ and those of $Q$ and $R$ is much less than $\pi/2$. This ensures
that $BD(P,R)$ is much larger than $BD(P,Q)$ and $BD(Q,R)$. Our result is
characterized by the following theorem.

\begin{theorem}
\label{lbembed-bhat}
There exist $d$-dimensional $\beta$-constrained distributions such that any
embedding of these distributions under the Bhattacharyya distance
measure into a metric space must incur a distortion of 
\[
D = \left\{ 
	\begin{array}{l l}
		\Omega \left(\frac{\ln \frac{1}{d\beta}}{\ln d}\right) & \quad \mbox{when $\beta > \frac{4}{d^2}$}\\
		\Omega \left(\frac{\ln \frac{1}{\beta}}{\ln d}\right) & \quad \mbox{when $\beta \leq \frac{4}{d^2}$}\\
	\end{array} \right.
\]
\end{theorem}
\begin{proof}
We know that the Bhattacharyya coefficient for two distributions is the
inner product between their corresponding vector representations on the unit
sphere. Recalling the definitions,
\begin{eqnarray*}
BC(p,q) &=& \sum_{i=1}^d \left( \sqrt{p_i}\cdot\sqrt{q_i} \right)\\
BD(p,q) &=& -\ln BC(p,q)
\end{eqnarray*}
Consider the following $\beta$-constrained distributions which satisfy the
above properties :
\begin{eqnarray*}
P &=& \left(1-(d-1)\beta, \beta , \dots ,\beta\right)\\
Q &=& \left(\frac{1}{d}, \frac{1}{d}, \dots ,\frac{1}{d}\right)\\
R &=& \left(\beta, 1-(d-1)\beta, \dots ,\beta\right)
\end{eqnarray*}
Now we have
\begin{eqnarray*}
\frac{BD(P,Q) + BD(Q,R)}{BD(P,R)} &=& \frac{\ln
\left({\left\langle{ \sqrt{P},\sqrt{Q} }\right\rangle  \cdot
\left\langle{ \sqrt{Q},\sqrt{R} }\right\rangle}\right)}{\ln
\left({\left\langle{ \sqrt{P},\sqrt{R}}\right\rangle}\right)}\\
\left\langle{ \sqrt{P},\sqrt{R} }\right\rangle &=& (d-2)\beta + 2
\sqrt{\beta - (d-1)\beta^2}\\
\left\langle{ \sqrt{P},\sqrt{Q} }\right\rangle &=& \frac{\sqrt{1-(d-1)\beta} + \sqrt{\beta}(d-1)}{\sqrt{d}}\\
&=& \left\langle{	\sqrt{Q},\sqrt{R} }\right\rangle
\end{eqnarray*}
Applying Lemma~\ref{lb} we get the following bound on the distortion
\[
D > \frac{\ln \left(\frac{1}{(d-2)\beta + 2 \sqrt{\beta -
(d-1)\beta^2}}\right)}{\ln \left(\frac{\sqrt{d}}{\sqrt{1-(d-1)\beta} +
\sqrt{\beta}(d-1)}\right)}
\]
Since $\beta >0$ and $1 - (d-1)\beta > \frac{1}{d}$, we have
$\sqrt{1-(d-1)\beta} + \sqrt{\beta}(d-1) > \frac{1}{d}$. In order to get a
lower bound on the numerator, we need an upper bound on $(d-2)\beta + 2
\sqrt{\beta}\sqrt{1 - (d-1)\beta}$. Using the fact that $1-(d-1)\beta \leq
1$, we get an upper bound of $d\beta + 2\sqrt{\beta}$ on the above
expression. In case $\beta > \frac{4}{d^2}$
\[
D = \Omega\left(\frac{\ln \frac{1}{d\beta}}{\ln d}\right)
\]
Otherwise, $d\beta + \sqrt{\beta} < 3\sqrt{\beta}$, which implies that
\[
D = \Omega\left(\frac{\ln \frac{1}{\beta}}{\ln d}\right)
\]
\end{proof}

In the following section, we demonstrate that this bound is tight up to a $O(d\ln d)$ factor.

\subsection{A Metric Embedding for the Bhattacharyya distance}
In this section, we first show that the Bhattacharyya distance is very closely
related to the Hellinger distance measure as previously defined. Since the
Hellinger distance forms a metric in the positive orthant, this allows us to get an upper bound on the
distortion which, for a fixed dimension, almost matches the lower bound.

\begin{theorem}
\label{membed-bhat}
For any two $d$-dimensional $\beta$-constrained distributions $P$ and $Q$
with $\beta < \frac{1}{2d}$, we have
\[
H(P,Q) \leq BD(P,Q) \leq \frac{d}{1-2\beta d} \ln
\frac{1}{(d-1)\beta}H(P,Q)
\]
\end{theorem}
\begin{proof}
For two distributions $P,Q$, recall 
\begin{eqnarray*}
BC(P,Q) &=& 1 - H(P,Q)\\ 
BD(P,Q) &=& -\ln \left(\sum_{i=1}^d \sqrt{p_i}\sqrt{q_i} \right)\\
  &=& -\ln \left(1 -H(P,Q)\right)\\
  &=& \sum_{k=1}^\infty \frac{H(P,Q)^k}{k}
\end{eqnarray*}
To arrive at the lower bound we truncate the infinite series at the first
term.	For the upper bound, we need to use the fact that the function $f(x)= -\ln
(1-x)$ is convex. The maximum Hellinger distance between any two
$\beta$-constrained distributions is $2(\sqrt{1-(d-1)\beta} - \beta)^2$.
Let,
$a = (\sqrt{1-(d-1)\beta} - \beta)^2$. Due
to convexity of $f$, the line $mx$ lies above the curve $-\ln (1-x)$ where $m
= \frac{f(a)}{a}$. Therefore, we have
\[
BD(P,Q)  = -\ln \left(1-H(P,Q)\right) \leq
\frac{1}{a}\ln\left(\frac{1}{1-a}\right) H(P,Q)
\]
Also, $1-a = (d-1)\beta + 2\beta\sqrt{1-(d-1)\beta} - \beta^2
\geq (d-1)\beta$ since $2\sqrt{1-(d-1)\beta} - \beta \geq 0$. Thus we get
\[
 BD(P,Q) \leq \frac{d}{1-2\beta d} \ln \left(\frac{1}{(d-1)\beta}\right)H(P,Q)
 \]
This implies that the identity embedding of a point set under the
Bhattacharyya distance into one under the Hellinger distance incurs a
distortion of $\frac{d}{1-2\beta d} \ln \frac{1}{(d-1)\beta}$.
\end{proof}

For constant $d$ and sufficiently small $\beta$, the lower bound presented in
Section~\ref{lb-bhat} is essentially $\Omega\left(\ln \frac{1}{\beta}\right)$,
whereas the embedding presented in this section has a distortion of
$O\left(\ln\frac{1}{\beta}\right)$ which implies that the lower bound is tight.
In general it can be seen using Theorems~\ref{lbembed-bhat} and \ref{membed-bhat} that for sufficiently small $\beta$ the lower bound presented is tight up to a factor of $O(d\ln d)$.
Further, the result presented in Theorem~\ref{hellinger-euclid} can be used to
perform dimensionality reduction as well.

\section{The Kullback-Leibler Divergence}
\label{kl}

The Kullback-Leibler divergence arises in information theoretic settings where
probability distributions are evaluated in terms of their entropy or the amount
of information they contain. The Kullback-Leibler divergence measures the
difference between the relative entropy and the self entropy of two
distributions. Given two histograms $P = \{p_1,p_2, \dots, p_d\}$ and $Q =
\{q_2,q_2 \dots q_d\}$, the Kullback-Leibler divergence between the two
distributions is defined as 
\[
KL(P,Q) =  \sum\limits_{i=1}^d p_i \ln \frac{p_i}{q_i}
\]
Informally, it gives an asymptotic lower bound on the overhead incurred in terms
of encoding length if one were to encode data assuming it to be generated by a
random source characterized by $Q$ when it fact the source is characterized by
$P$. The Kullback-Leibler divergence is non-symmetric and unbounded, i.e., for
any given $c > 0$, one can construct histograms whose Kullback-Leibler divergence
exceeds $c$. In order to avoid these singularities, we assume that the
histograms are $\beta$-constrained.

\begin{lemma}
\label{kl-bounds}
Given two $\beta$-constrained histograms $P$, $Q$, $ 0 \leq KL(P,Q) \leq
\ln\frac{1}{\beta}$. 
\end{lemma}
\begin{proof}
The lower bound follows directly from Jensen inequality \cite{dhs}. For the upper bound, since we know that $\frac{p_i}{q_i} \leq \frac{1}{\beta} $ for all $i
= 1,2, \dots, d$, we can write 
\[
KL(P,Q) =  \sum_{i=1}^d p_i \ln \frac{p_i}{q_i}  \leq  \sum_{i=1}^d p_i
\ln \frac{1}{\beta} = \ln \frac{1}{\beta}
\]
For $\beta = O\left(\frac{1}{d}\right)$, the upper bound
is tight up to a	constant factor.
\hfill{}
\end{proof}

In the following discussion, we show that low distortion embeddings into metric
spaces cannot exist for the Kullback-Leibler divergence. In order to show this,
we utilize the relaxed triangle inequality presented in Section~\ref{rti} and
explicitly construct point sets that violate the relaxed triangle inequality. We
also develop another technique that exploits the fact that the Kullback-Leibler
divergence is not symmetric. We first present this new proof technique before
moving on to prove bounds utilizing both the proof techniques.

\subsection{The Asymmetry Technique}

We present a general result that can be used to prove lower bounds on the
embedding distortion when we intend to embed a non-symmetric distance measure
into a metric space. The idea is to exploit the existence of two points $p,q$
for which there is a large gap between the distances between $p$ to $q$ and $q$
to $p$. This idea is formalized in Lemma~\ref{asym-lb} using the following definition.
\begin{definition}[$\gamma$-Relaxed Symmetry]
A set $X$ equipped with a distance function $d:X \times X \longrightarrow
\mathbb{R}^+\cup\{0\}$, is said to satisfy \emph{$\gamma$-relaxed symmetry} if
there exists $\gamma \geq 0$ such that for all point pairs $p,q \in X$, the following holds
\[
|d(p,q) - d(q,p)| \leq \gamma
\]
Note that metrics satisfy the $\gamma$-relaxed symmetry for $\gamma = 0$.
\end{definition}
\begin{lemma}
\label{asym-lb}
Given a set $X$ equipped with a distance function $d$ that does not satisfy
the $\gamma$-relaxed symmetry such that $d(x,y) \leq M$ for all $x,y \in X$,
any embedding of $X$ into a metric space incurs a distortion of at least
$1 + \frac{\gamma}{M}$.
\end{lemma}
\begin{proof}
Since $(X,d)$ does not satisfy the $\gamma$-relaxed symmetry, there exist
points $p,q \in X$ such that $|d(p,q) - d(q,p)| > \gamma$. If $(X,d)$ is
embeddable via a mapping $f$ into a metric space $(Y, \rho)$ with a
distortion of $D$ then for some constant $r > 0$, 
\[
r\cdot d(x,y) \leq \rho\left( {f(x),f(y)} \right)
\leq D \cdot r \cdot d(x,y)
\]
Without loss of generality, assume that $d(p,q) > d(q,p)$. Since $(Y, \rho)$
is a metric space, it	is symmetric, which implies that $\rho(f(p),f(q)) = \rho(f(q),f(p))$. Since
$d(p,q) > d(q,p) + \gamma$, we have
\begin{eqnarray*}
\rho(f(p), f(q)) - r\cdot d(q,p) &>& r \cdot \gamma\\
\frac{\rho(f(p), f(p))}{r\cdot d(q,p)} - 1 &>&  \frac{r	\cdot \gamma}{r\cdot d(p,q)}\\
D \geq \frac{\rho(f(p), f(p))}{r\cdot d(q,p)} &>& 1 +	\frac{\gamma}{M}
\end{eqnarray*}
This implies that the distortion is at least $1 + \frac{\gamma}{M}$.
\end{proof}

\subsection{Lower Bounds on Distortion for Embeddings into Metric Spaces}

We now apply the above lemma to show that one cannot hope to obtain an almost isometric embedding of the Kullback-Leibler divergence into any metric space. We show the existence of two histograms $P$ and $Q$ such that $|KL(P,Q)-KL(Q,P)|$ is large. The result is formally stated in the following theorem. 

\begin{theorem}
For sufficiently large $d$ and small $\beta$, there exists a set $S$
of $d$-dimensional $\beta$-constrained histograms and a constant $c > 0$
such that any embedding of $S$ into a metric space incurs a distortion of at
least $1 + c$.
\end{theorem}
\begin{proof}
The idea is to choose two distributions which violate a $\delta$-relaxed symmetry for large $\delta$.
Consider the following two distributions where $\beta \neq \frac{1}{d}$
\begin{eqnarray*}
P &=& \left\{\frac{1}{d},\frac{1}{d}, \dots, \frac{1}{d}\right\}\\
Q &=& \{1-(d-1)\beta,\beta,\ldots,\beta\}
\end{eqnarray*}
Define $\Delta KL(P,Q) = |KL(P,Q) - KL(Q,P)|$. Since we have
\begin{eqnarray*}
KL(P,Q) &=& \left( 1- \frac{1}{d}\right) \ln \frac{1}{\beta d} - \frac{1}{d}
\ln (d(1-(d-1)\beta))\\
KL(Q,P) &=& \beta (d-1) \ln \beta d + (1-(d-1)\beta) \ln (d(1-(d-1)\beta))
\end{eqnarray*}
by rearranging the terms we get 
\[
\Delta KL(P,Q) = \left |\left(1-\frac{1}{d}+\beta(d-1)\right)\ln\frac{1}{\beta d} -  \left(\frac{1}{d} + 1 - (d-1)\beta\right) \ln(d(1 -(d-1)\beta))
\right|
\]
For sufficiently large $d$, we consider the situation when $\beta$ is large
i.e. say $\beta = \frac{1}{\Theta(d)}$. In this case we have $\Delta KL(P,Q)
= \left| \Theta\left(\ln \frac{1}{d\beta}\right) - \Theta(\ln d) \right| =
\Theta(\ln d)$. Since for $\beta$-constrained distributions, the maximum
inter point distance i.e. $M = \ln\frac{1}{\beta}$, we get the a lower bound
on the distortion by using Lemma~\ref{asym-lb} as $1 + \frac{\Theta(\ln
d)}{\ln\frac{1}{\beta}}$ which is $1 + \Theta(1)$ as $\beta =
\frac{1}{\Theta(d)}$.

For small $\beta$ - say $\beta = o\left(\frac{1}{d^4}\right)$, since $d > 2$ we get
\[
\Delta KL(P,Q) \geq \frac{1}{2}\ln\frac{1}{\beta d} - \frac{3}{2}\ln d = \frac{1}{2}\ln\frac{1}{\beta d^4}
\]
Using a similar argument as above, we get a lower bound on the distortion as
\[
D = 1 + \Omega\left(\frac{\ln\frac{1}{\beta d^4}}{\ln\frac{1}{\beta}}\right) = 1 + \Omega(1)
\]
Hence we conclude that any embedding of point sets which contain the points $P$ and $Q$ into a metric
space must have a distortion $D$ of at least $1 + c$ for some constant $c$.
\end{proof}

The above argument shows the impossibility of obtaining almost isometric (i.e.,
with distortion arbitrarily closed to 1) embeddings of $\beta$-constrained
histograms under the Kullback-Leibler divergence. Since $\Delta (P,Q)$ can be at
most $\ln \frac{1}{\beta}$, for any two $\beta$-constrained distributions, using
this technique, a different choice of points can at best provide a constant
factor improvement over the above bounds. It should be noted that one
cannot hope to get significant improvement on the above bounds via this
technique by choosing two different points in the probability simplex. However,
an application of the relaxed triangle inequality technique shows that the
situation is much worse. We will now demonstrate that the Kullback-Leibler
divergence admits point sets which violate the $\lambda$-relaxed triangle
inequality, where $\lambda$ can be made arbitrarily small. This implies (by
applying Lemma~\ref{lb}) that the distortion into any metric space can be made
arbitrarily large.

\begin{theorem}
For sufficiently large $d$, there exist $d$-dimensional $\beta$-constrained
distributions such that embedding these under the Kullback-Leibler
divergence into a metric space must incur a distortion of $\Omega
\left(\frac{\ln \frac{1}{d\beta}}{\ln \left(d\ln
\frac{1}{\beta}\right)}\right)$.
\end{theorem}
\begin{proof}
We construct three $\beta$-constrained distributions that fail
to satisfy a relaxed triangle inequality under the Kullback-Leibler divergence.
Consider the following distributions. The parameters $\epsilon$
and $c$ will be fixed later. Let
\begin{eqnarray*}
P &=& \left(\frac{1}{d}, \frac{1}{d}, \dots ,\frac{1}{d}\right)\\
Q &=& \left(1-(d-1)\epsilon, \epsilon , \dots ,\epsilon\right)\\
R &=& \left(1-(d-1)e^{-c}, e^{-c} , \dots ,e^{-c}\right)
\end{eqnarray*}
where $\frac{1}{d} \geq \epsilon > e^{-c} \geq \beta $.  We have
\begin{eqnarray*}
KL(P,Q) &=& \left(1- \frac{1}{d} \right)\ln \frac{1}{d\epsilon} +
\frac{1}{d} \ln \frac{1}{d(1-(d-1)\epsilon)}\\
        &\leq& \ln \frac{1}{d\epsilon}\\
KL(Q,R) &=& (1-(d-1)\epsilon) \ln \frac{1-(d-1)\epsilon}{1-(d-1)e^{-c}} + (d-1) \epsilon \ln (\epsilon e^c)\\
        &\leq& (d-1) \epsilon \ln \epsilon  + (d-1) c\epsilon\\
KL(P,R) &=& \left(1- \frac{1}{d} \right)\ln \frac{1}{de^{-c}} + \frac{1}{d}
\ln \frac{1}{d(1-(d-1)e^{-c})}\\
        &\geq& \frac{1}{2}(c - \ln d) + \frac{1}{d}\ln \frac{1}{d}\\
        &=& \Omega(c - \ln d) - O(1)
\end{eqnarray*}
Using the above inequalities,
\begin{eqnarray*}
\lambda &=& \frac{KL(P,Q) + KL(Q,R)}{KL(P,R)}\\
        &=& O\left(\frac{ \ln \frac{1}{d\epsilon} + (d-1) \epsilon \ln \epsilon  + (d-1) c\epsilon}{c - \ln d}\right)
\end{eqnarray*}
Hence, any point set containing these three points violates the
$\lambda$-relaxed triangle inequality. Now, using Lemma \ref{lb}, the
distortion for the Kullback-Leibler divergence is
\[
D > \frac{1}{\lambda} = \Omega\left(\frac{c - \ln d}{ \ln
\frac{1}{d\epsilon} + d \epsilon \ln \epsilon  + dc\epsilon}\right)
\]
Since $\epsilon\ln\epsilon < 0$, hence
\[
D = \Omega\left(\frac{c - \ln d}{ \ln
\frac{1}{d\epsilon} + dc\epsilon}\right)
\]
Consider the function $f(c,\epsilon) = \frac{c - \ln d}{ \ln
\frac{1}{d\epsilon} + dc\epsilon}$. It turns out that $\frac{\partial
f}{\partial c} > 0$ for all values of $c$. Hence, the maxima is achieved at
the maximum value of $c$ which is $\ln \frac{1}{\beta}$. Furthermore we find
that $\frac{\partial f}{\partial\epsilon} = 0$ at $\epsilon = \frac{1}{dc} =
\frac{1}{d\ln \frac{1}{\beta}}$. It can be confirmed that this extrema is
actually a maxima. For a fixed value of $d$ we can choose $\beta$ small
enough to make sure that the value of $\epsilon$ is at least $\beta$. For
these values of $c$ and $\epsilon$ we get the lower bound as
\[
D = \Omega \left(\frac{\ln \frac{1}{d\beta}}{\ln \left(d\ln
\frac{1}{\beta}\right) +  1}\right)
\]
Thus, the result follows.
\end{proof}

\noindent
\textbf{Interpreting the Lower Bounds}: The above bounds show how the
Kullback-Leibler divergence behaves near the uniform distribution and near the
boundaries of the probability simplex. The bounds indicate that near the uniform
distribution, asymmetry makes the Kullback-Leibler divergence hard to
approximate by a metric but as we move away from the uniform distribution the
hardness is because of the violation of the relaxed triangle inequality. More
formally, it can be seen that for point sets which are $\beta$-constrained for
large $\beta$ (say $\beta = \Omega\left(\frac{1}{d}\right)$), the lower bound
using the asymmetry argument gives a $1 + \Theta(1)$ bound whereas the triangle
inequality argument gives a $o(1)$ bound. For smaller $\beta$ (say $\beta =
o\left(\frac{1}{d^4}\right)$) we get a better lower bound using the relaxed
triangle inequality argument. This lower bound behaves asymptotically as
$\frac{\ln \frac{1}{\beta}}{\ln\ln\frac{1}{\beta}}$ which can be made
arbitrarily large. Note that the asymmetry argument gives a constant lower
bound even for very small $\beta$ but this only shows the ineffectiveness of
this proof technique for small $\beta$.

\subsection{An Embedding for the Kullback-Leibler divergence}

In this section, we examine the properties of the identity embedding of point
sets under the Kullback-Leibler divergence into the $l_2^2$ distance measure.
Once we have obtained this result one can again apply JL-type embeddings to achieve
dimensionality reduction as well. To obtain this bound we use a well known
inequality in information theory due to Pinsker \cite{top-id}. Our result is
characterized by the following theorem.

\begin{theorem}
For any two $d$-dimensional $\beta$-constrained distributions $P$ and $Q$,
\[
\frac{l_2^2(P,Q)}{2} \leq KL(P,Q) \leq \left(\frac{1}{2\beta} +
\frac{1}{3\beta^5}\right)l_2^2(P,Q)
\]
\end{theorem}
\begin{proof}
The lower bound essentially follows from Pinsker's inequality \cite{top-id} which states
that
\[
\frac{l_1^2(P,Q)}{2} \leq KL(P,Q)
\]
Since $l_1(P,Q) \geq l_2(P,Q)$, the lower bound follows automatically. For the upper bound, we use Taylor's Theorem on the expression for $KL(P,Q)$
\begin{eqnarray*}
KL(P,Q) &=& \sum\limits_{i=1}^n p_i \ln \frac{p_i}{q_i}\\
        &=& \sum\limits_{i=1}^d -p_i \ln \left(1 - \frac{p_i-q_i}{p_i}\right)
\end{eqnarray*}
By Taylor's Theorem, there exists $\epsilon$ such that $|\epsilon| \leq
\left|\frac{p_i-q_i}{p_i}\right|$, for which the following holds :
\begin{eqnarray*}
KL(P,Q) &=& \sum\limits_{i=1}^d p_i \left(\frac{p_i-q_i}{p_i}+
\frac{(p_i-q_i)^2}{2p_i^2} +
\frac{1}{(1-\epsilon)^3}\cdot\frac{(p_i-q_i)^3}{3p_i^3}\right)\\
        &=& \sum\limits_{i=1}^d \left(\frac{(p_i-q_i)^2}{2p_i} +
\frac{1}{(1-\epsilon)^3}\cdot\frac{(p_i-q_i)^3}{3p_i^2}\right)
\end{eqnarray*}
Since $|\epsilon| \leq \left|\frac{p_i-q_i}{p_i}\right|,
\frac{1}{(1-\epsilon)^3} \leq \frac{1}{\beta^3}$, and
$\left|\frac{p_i-q_i}{p_i}\right| \leq \frac{1}{\beta}$, we have
$\frac{1}{(1-\epsilon)^3}\cdot \frac{(p_i-q_i)^2}{3p_i}\cdot
\frac{(p_i-q_i)}{p_i} \leq \frac{(p_i-q_i)^2}{3\beta^5}$. Therefore,
\[
KL(P,Q) \leq \left(\frac{1}{2\beta} + \frac{1}{3\beta^5}\right)l_2^2(P,Q)
\]
The distortion of this identity embedding is $O\left(\frac{1}{\beta^5}\right)$.
\end{proof}
Although this embedding does not give us a provably small distortion, it
can still be used in practical situations since the embedding is into $l_2^2$ space
and it allows for low distortion dimensionality reduction using the
Johnson-Lindenstrauss Lemma.

\section{The Class of Quadratic Form Distance Measures}
\label{qfd}

Consider the vector space $\mathbb{R}^d$. Given a $d \times d$ positive definite
matrix $A$, the \emph{Quadratic Form Distance measures (QFDs)} define a distance
measure over $\mathbb{R}^d$. If $x, y \in \mathbb{R}^d$, then $Q_{A}(x, y)$ is
defined to be 
\[
Q_{A}(x, y) = \sqrt{(x-y)^{T}A(x-y)}
\]
These distance measures can be seen as acting on a distorted Euclidean space.
When $A$ is a diagonal matrix with positive entries, the corresponding QFDs are
weighted Euclidean distance measures. The family of quadratic form distances are
actually defined for general positive semi-definite matrices. However, for
positive definite matrices, the distance measure $Q_{A}$ forms a metric. We now
show that every QFD can be embedded into a low-dimensional space with low
distortion in the inter-point distances.

\begin{theorem}
\label{qf-euclid}
The family of quadratic form distance measures admit a low distortion
JL-type embedding.
\end{theorem}
\begin{proof}
Every quadratic form distance measure forming a metric is characterized by a
square matrix $A$ which is positive definite. However, every positive
definite matrix $A$ can be subjected to a Cholesky Decomposition of the form
$A = L^TL$ \cite{cholesky}. Consider the transformation $x \longmapsto
R\left( {Lx} \right)$ where $R$ is the random projection matrix involved in
the Johnson-Lindenstrauss Lemma.  Consider two points $x,y \in
\mathbb{R}^d$, then 
\[
Q_A(x,y) = \sqrt{(x-y)^{T}A(x-y)} = \sqrt{(L(x-y))^{T}(L(x-y))}
\]
which is the Euclidean distance between the points $Lx$ and $Ly$. Thus, the
proposed transformation gives us a low distortion embedding since the
problem has been reduced to that in the undistorted Euclidean space where
the Johnson-Lindenstrauss Lemma is applicable.
\end{proof}

The previous theorem can be easily seen to provide a simple algorithm to reduce the
dimensionality of the data points and still preserve the distance as given by the QFD.
Given this formulation we now look at the Mahalanobis distance which is a special case
of the QFD measure where the positive definite matrix is taken to be the covariance matrix
of a normal multivariate probability distribution. Given the construction of low distortion
embeddings for QFDs, the following result is immediate.
\begin{corollary}
The Mahalanobis metric admits a randomized polynomial time low distortion
JL-type embedding.
\end{corollary}

\section{Conclusions}
\label{conclude}

We have investigated various statistical distance measures from the point of view of dimensionality reduction and embeddability into metric spaces. We examined and presented novel dimensionality reduction techniques for the Bhattacharyya distance, the Hellinger distance and the Mahalanobis
distance measure using the Johnson-Lindenstrauss Lemma.

We also examined the question of finding low distortion embeddings of the Bhattacharyya distance and the Kullback-Leibler divergence which are non-metric distance measures into metric spaces. We developed two novel techniques that can be used to prove lower bounds on the distortion that must be incurred by any such embedding.

For the Bhattacharyya distance, we demonstrated that the lower bound presented is almost tight by analyzing its relationship between the Hellinger distance which forms a metric. We performed a similar exercise for the Kullback-Leibler divergence to relate it with the $l_2^2$ measure. Although
it does not match the lower bounds, it is of practical significance since it allows for dimensionality reduction.

The question that we leave open is that of dimensionality reduction under the Kullback-Leibler divergence. Our preliminary investigations show that this is unlikely if one wants the resulting embedded objects to lie on the low-dimensional probability simplex. Also, it remains to be seen if the methods developed in this paper, viz., the adaptations of random projections to various distance measures and the proof techniques developed, find applications for other widely used statistical distance measures, many of which are non-metrics.
\bibliography{ref}

\end{document}